\documentclass[prb,aps,preprint,dcolumn,superscriptaddress,showpacs,amsmath,amssymb,showkeys,floatfix]{revtex4} 
\usepackage{graphicx}
\begin{document}
\title{Time relaxation of interacting single--molecule magnets}
\author{Julio F. Fern\'andez}
\affiliation{ICMA, CSIC and Universidad de Zaragoza, 50009-Zaragoza, Spain}
\email[E-mail address: ] {Jefe@Unizar.Es}
\homepage[ URL: ] {http://Pipe.Unizar.Es/~jff}
\author{Juan J. Alonso}
\affiliation{F\'{\i}sica Aplicada I, Universidad de M\'alaga,
29071-M\'alaga, Spain}
\email[E-mail address: ] {jjalonso@Darnitsa.Cie.Uma.Es}
\date{\today}

\begin{abstract}
We study the relaxation of 
interacting single--molecule magnets (SMMs)
in both spatially ordered and disordered
systems. The tunneling window is assumed to be, as in Fe$_8$,
much narrower than the dipolar field spread.
We show that relaxation in disordered systems
differs qualitatively from relaxation
in fully occupied cubic and Fe$_8$ lattices.
We also study how line shapes 
that develop in ``hole--digging''
experiments evolve with time $t$ in these fully occupied
lattices. We show (1) that the dipolar field $h$
scales as $t^p$ in these 
hole line shapes and show (2) how $p$ varies 
with lattice structure. Line shapes are not,
in general, Lorentzian. More specifically, in the lower
portion of the hole, they behave as $(\mid h\mid /t^p)^{(1/p)-1}$
if $h$ is outside the tunnel window. This
is in agreement with experiment and with our own 
Monte Carlo results.
\vspace{1pc}
\end{abstract}

\maketitle
\maketitle
\section{introduction}
Magnetic relaxation in crystals of single--molecule magnets (SMM's), such as 
Fe$_8$, has become a
subject of great interest.\cite{MQTe0,fried,sanww,sc}
At low temperature, $T$, each SMM
behaves approximately as a single spin $S$.
Magnetic relaxation at 
$k_BT\lesssim 0.1U/S$,
where $U$ is a magnetocrystalline anisotropy barrier, 
is temperature 
independent, and is duly
attributed to quantum tunneling under the barrier. 
Hyperfine interactions with nuclear spins 
as well as dipole--dipole interactions among all SMM electronic spins 
give rise to a variety of phenomena that are not yet fully understood.
Hyperfine interactions enable spins to tunnel even when the ensuing
Zeeman energy change $2\varepsilon_h$ is much larger than the tunnel 
splitting energy $\Delta$, provided $\mid \varepsilon_h \mid$ is smaller
than some $\varepsilon_w$.\cite{PS} For Fe$_8$, for instance,
$\Delta\sim 10^{-4}$ mK, $\varepsilon_w\approx 10$ mK, and the rms value
of the Zeeman energy $\delta \varepsilon_h$ is approximately 400 mK. 
We shall restrict
ourselves to systems with 
$\varepsilon_w\ll\delta\varepsilon_h$.
In these systems, the relaxation of the magnetization goes
on long after time $\Gamma^{-1}$, where $\Gamma$
is the spin tunneling rate for spins 
within the tunneling energy window
(that is, spins for which 
$\mid \varepsilon_h \mid <\varepsilon_w$).
Tunneling spins give rise to changing dipolar fields,
which in turn bring new spins into the tunneling energy window,
thus keeping a magnetic relaxation process from extinction.
In this paper, we focus our attention on effects that stem
from this process.

We consider here experiments of the sort that were
reported by Wernsdorfer et al. in Ref. [\onlinecite{ww}].
In them, a crystalline sample of SMM's is 
first quenched from $k_BT\sim U/S$ 
to $k_BT\lesssim 0.1U/S$
in either (a) a weak applied magnetic field
of a few mT, and later observed after the
field is switched off at time $t=0$ or (b) 
a zero field, and later observed after a weak field
is applied at $t=0$. We shall refer to the former
as a FC (for field cooled) experiment  and to the latter 
as a ZFC (for zero field cooled) experiment.
Both types of experiments can be lumped into
one defining
$\tilde m\equiv 1-m/m_0$, where $m_0$ is the initial
magnetization in a FC experiment,
and $\tilde m\equiv m/m_s$,
where $m_s$ is the final steady state magnetization
in a ZFC experiment.
Wernsdorfer et al.
observed the
$\tilde m\propto t^{1/2}$ over
roughly two time decades. 
The time evolution of ``holes'' that, under suitable conditions, develop
in a magnetization density function, have also been reported.\cite{ww,wwq}
The existing theory at the time\cite{PS,villa,chudno} 
predicted a universal $\sqrt{t}$ short time relaxation
from a fully polarized system,
but said little about relaxation from or into weakly polarized
states.\cite{PS2} Other theories\cite{miya} take hyperfine interactions
into account but disregard dipole--dipole interactions.
They therefore apply if
$\delta \varepsilon_h \lesssim\varepsilon_w$
which is not within the scope of this paper. 

We have developed a theory\cite{Let,ourPRB} that gives the time
evolution both of $\tilde m$ and of the holes' line shapes
in weakly polarized systems of interacting SMMs, such as Fe$_8$,
in which $\varepsilon_w\ll\delta\varepsilon_h$.
There are three clearly discernible time regimes.
For $\Gamma t \lesssim 1$, $\tilde m\propto \Gamma t$. 
In the second time stage,
when $1\lesssim \Gamma t$, up to
some time before $\tilde m\sim 1$,
$\tilde m\propto t^p$, at least for all fully occupied cubic
lattices. 
Moreover, the theory gives a simple relation
that specifies how $p$ varies with lattice structure.
In FC experiments, $\tilde m\sim 1$ in the third time stage,
that is, $m\approx 0$. The third time stage
is more interesting in ZFC experiments.
Then $m(t)$ settles down
temporarily to a quasi stationary
value $m_s$, which the theory predicts,
if either $k_BT\gg \varepsilon_w$ or
if the heat exchange rate with the lattice is much smaller
than $\Gamma$; 
on the other hand, if $k_BT\gg \varepsilon_w$ is not fulfilled
and heat exchange rate with the lattice is not much smaller
than $\Gamma$, then the relaxation of the magnetization
shifts into a thermally driven approach to equilibrium,
skipping the quasi stationary state.
A quite different treatment of relaxation
from weakly polarized states that gives
$1-m/m_0\propto \sqrt{t}$,
independently of the spins' spatial distribution,
is given by Tupitsyn, Stamp, and
Prokof'ev (TSP) in Ref. [\onlinecite{TSP}], 
criticized in Ref. [\onlinecite{comm}],
and defended in Ref. [\onlinecite{TSP2}]. 
TSP's treatment of relaxation from {\it weakly}
polarized states is rather unrelated to the
earlier theory\cite{PS} of Prokofe'v and Stamp for
a $\sqrt{t}$ relaxation from {\it fully} polarized
states, about which we have nothing to say here.
According to TSP, $\sqrt{t}$ relaxation 
in weakly polarized systems holds
as long as $1\lesssim \Gamma t$ and $\tilde m\ll1$,
that is, roughly, over the second time stage. This is also
the time domain where our theory gives 
$\tilde m\simeq (\varepsilon_w/\delta\varepsilon_h)
(\Gamma t)^p$ for fully occupied lattices.
This time stage, sometimes referred to as ``short times'',
can in fact be arbitrarily large, for arbitrarily small
$\varepsilon_w$, since $\tilde m\ll 1$ only implies 
$\Gamma t\ll (\delta\varepsilon_h/\varepsilon_w)^{1/p}$.
Note also that
$\sqrt{t}$ relaxation from {\it fully} polarized systems
ends sooner, when 
$t \sim \delta\varepsilon_h/\varepsilon_w$.
\cite{PS} 
This may explain why Wernsdorfer\cite{ww} et al. 
were able to observe $\tilde m\sim \sqrt{t}$ in Fe$_8$ 
($p\approx 0.5$ for Fe$_8$) under
weak applied magnetic fields for up to some $10^3$ s
while Ohm\cite{ohm} et al. 
observed $\sqrt{t}$ relaxation 
from fully polarized Fe$_8$
for only up to $10^2$ s.

The two approaches, ours and TSP's, are quite different.
The underlying assumptions
are as follows. In Ref. [\onlinecite{TSP}], the main result
follows from a claim that is made on
the function
$f(h ,t)=p_\downarrow (h ,t)-p_\uparrow (h ,t)$,
where $p_\uparrow (h ,t)$ and $p_\downarrow (h ,t)$ are the
number densities of up-- and
down--spins, respectively, with a dipolar
field $h$ acting on them at time $t$.
In its latest form,\cite{TSP2}
the claim is that the scaling form $f(h ,t)=f(h/t^2)$
``was found to be valid in our MC simulations 
for different lattice types...'' when $1\lesssim \Gamma t$ and
$\tilde m\ll 1$.
Since, the magnetization $m$ and $f(h,t)$
are clearly related by $m(t)=-\int dh f(h,t)$,
a $\sqrt{t}$ relaxation follows immediately.
As we have shown recently,\cite{comm} the above scaling form
holds approximately for SC and Fe$_8$ lattices (as defined in
Ref. [\onlinecite{Fel}]),
but not in general. It fails, for example, for
FCC, BCC, and diamond lattices.\cite{more}
Our theory also gives the time
evolution of $f(h,t)$, but follows from 
a more fundamental assumption: that
the dipolar field on any one given site 
changes by some {\it random} amount $\Delta h$,
whenever a spin flips somewhere else for the first time, and that
$\Delta h$ follows a Lorentzian distribution
(for more details see
Sects. \ref{rtr} and \ref{th}, and Ref. [\onlinecite{ourPRB}]).
The integro-differential
equations for the evolution of $f(h,t)$ and of the
magnetization that obtain in our theory, follow from this assumption.

Unfortunately, as far as we know, only experiments on
a crystalline Fe$_8$ structure have thus far
been performed. 
However, MC simulations have been performed for various fully occupied
cubic lattices, which have given values of
$p$ that agree with our predictions.\cite{ourPRB,comm}
This paper's first
aim is to extract from our theory how the magnetization
is supposed to relax in Fe$_8$, compare this with experiment
over the time span where published experimental data exists,\cite{ww}
and make predictions for later times.
It is also our purpose to predict, and check with MC simulations,
how $m$ relaxes with $t$ for other spatial distributions,
namely, under full spatial disorder, thus providing another test 
for our theory.

The holes observed in the experiments described above\cite{ww}
are also of interest. They
correspond to ``wells'' that develop in the function
$f(h ,t)=p_\downarrow (h ,t)-p_\uparrow (h ,t)$ (defined above).
From the relation $m(t)=-\int dh f(h,t)$, the time evolution
of $m$ follows,
but  $f(h,t)$ provides
additional information
about the magnetic evolution of the system that $m(t)$ does not.\cite{info}
For short times, that is, for
$\Gamma t \lesssim 1$,
the hole's width is equal to $\varepsilon_w$.\cite{JMM}
This was first surmised by Wernsdorfer et al\cite{ww}
to propose a number, approximately $10$ mK, for the
tunneling energy window, $\varepsilon_w$.
However, we know of no published data for the holes' line shape
evolution well into the intermediate time range,
that is, for $1\ll \Gamma t $. Our second aim
is to fill this gap. To this end, we work out from our theory
the time evolution of holes' line shapes 
in this time stage in fully occupied cubic systems
and Fe$_8$ crystals, and check the results obtained
against our MC results. We also obtain 
new results
of a more general nature for the holes' line shape. 
Before we state our results, we specify the model.

\subsection{The model}
\label{rt}

All spins are on a lattice,
they point along the easy
anisotropy axis, and interact as magnetic dipoles.
We consider both fully and partially occupied lattices.
Let the magnetic field at site $i$ produced by spin $S_j$
at site $j$ be given, in the usual notation,
by $h_{ij}=h_d(a/r_{ij})^3[1-3(z_{ij}/r_{ij})^2]$,
where $r_{ij}$ is the distance between the
$i$ and $j$ sites, $a$ is the distance between nearest neighbor sites,
$h_d=(\mu_0/4\pi )g\mu_BS/a^3$, and $S_i=\pm S$ for all $i$.
Furthermore, let the magnetic field $h_i$ at site $i$
be given by $h_i=\sum_j h_{ij}$, where $\sum_j$ is over
all occupied lattice sites. The tunnel 
window size and tunneling rate $\Gamma$
are defined next. At very low temperature,
that is, if $k_BT\lesssim 0.1U/S$, a spin can flip
only if the field $h$ acting on it satisfies
$\mid h\mid <h_w$.\cite{model}
The flipping rate is $\Gamma$ if upon tunneling
the energy decreases, but if the energy increases by
$\Delta E$, then the rate is $\Gamma \exp (-\Delta E/k_BT)$,
following detailed balance. (Even though 
$\mid \Delta E\mid <\varepsilon_w$, and ususally $k_BT\gg \varepsilon_w$,
$\exp (-\Delta E/k_BT)$ is not quite equal to $1$, and this makes
a difference after a sufficiently long time, as is shown below.)
We also simulate relaxation processes
in which the energy $E$ is assumed to remain constant (such
as if no spin-lattice relaxation takes place).
Then, we assume a value
of $T$ is such that $E$ remains approximately constant.
No tunneling window restriction
applies for spin flips if $k_BT\gtrsim U/S$.

We let $p(h,t)$ be 
the probability density function (PDF) that any one given 
spin have field $h$ at time $t$, let
$p_0(h)$ be the same distribution for
a completely random spin configuration,
and let 
\begin{equation}
\sigma =[\sqrt{2\pi}p_0(0)]^{-1}.
\end{equation}
For a Gaussian field distribution,
$\sigma$ is equal to the dipole field rms value
$\delta h$ for a random spin configuration (see Table I), 
but this is not so in general. Values of $\sigma$ 
that follow from MC simulations for
cubic and Fe$_8$ lattices with randomly oriented spins
are given in Table I. 
Finally, let
$h_0=(8\pi^2/3^{5/2})h_d\tilde n$, where $\tilde n$ is the
number of dipoles per unit cubic cell.\cite{ncc}
Randomly oriented spins on a cubic lattice give
a Lorentzian field distribution of $h_0$ 
half width at half maximum
if $\tilde n\ll 1$.\cite{abrag}
Values we will be using for $h_0$ are given in Table I.
From here on, unless otherwise stated, all magnetic fields and
energies are given in terms
$h_d$ and $g\mu_Bh_dS$, respectively.

\begin{table}
\caption{\label{tabla}Quantities $h_0$,
$\sigma$, and $\sqrt{\langle h^2\rangle_0}$ are given 
for randomly oriented
spins on the lattices specified. 
$h_0=\tilde n 8\pi^2/3^{5/2}$, $\sigma\equiv
[\sqrt{2\pi}p(0)]^{-1}$, where
$p(0)$ stands for the PDF at $h=0$,
and $\sqrt{\langle h^2\rangle_0}$ is the rms spatial 
average of the dipolar field.
For all lattices except for Fe$_8$, $h_0$, $\sigma$, and
$\sqrt{\langle h^2\rangle_0}$ are given in terms of $h_d$.
All lattices are fully occupied, except for
the ``dilute SC'', which stands for a SC lattice
with $\tilde n\ll 1$ sites occuppied. Finally, 
$\alpha=8\pi^2/3^{5/2}$.}
\begin{ruledtabular}
\begin{tabular}{|c|l|l|l|||}
LATTICE        &$h_0$          & $\sqrt{\langle h^2\rangle_0} $
&$\sigma$  \\ \colrule
SC & $\alpha$  &3.655    & 3.83(2)        \\
BCC & 2$\alpha$  & 3.864  & 4.03(2)         \\
FCC & 4$\alpha$  & 8.303  & 8.44(2)        \\
Fe$_8$\footnotemark[2] & 47(1) mT & 46(1) mT & 31(1) mT \\
dilute SC & $\alpha\tilde n$ &  & $\sqrt{\pi /2}h_0$     \\
\end{tabular}
\end{ruledtabular}
\footnotemark[2] We assume an easy anisotropy axis as given in
Refs. [\onlinecite{JF,barra,err}]
\end{table}

\subsection{Plan and main results}

Section \ref{rm} is devoted to the
relaxation of the magnetization. The equations from our 
theory\cite{ourPRB} which we use to
calculate the time evolution of the magnetization are restated in
Sect. \ref{rtr}. The results that follow from them for Fe$_8$
crystals are shown to agree rather well
with experiment and with our own MC results
in Sect. \ref{1cwe}. The evolution we predict for
$m(t)$ as well as our MC results cover a time
span that is 2 orders of magnitude longer than
the the experimental one $\tau_E$ ($\tau_E\approx 20$ minutes). 
Let $\tau_w$ be the end time for the regime where $m\propto t^p$.
We showed in in Ref. [\onlinecite{ourPRB}] that
$\tau_w\approx \Gamma^{-1}(\sigma /h_w )^{1/p}$, from
which we obtain $\tau_w\sim 10 \tau_E$.
For $\tau_w\lesssim t$, the evolution of $m(t)$ is shown to
depend sensitively on $T$ if good thermal contact
with a heat reservoir is assumed. If on the other 
hand we assume constant energy processes
(i.e., no spin-lattice relaxation), then $m$ levels off,
if only temporarily, to a stationary value $m_s$ when $\tau_w\lesssim t$.
The value of $m_s$ we obtain from
theory is unrelated to the equilibrium value of $m$, which only
obtains much later. This stage, when $m\rightarrow m_s$, sets in after
most spins in the system have tunneled at least once after
the magnetic field is applied. Simulations bear this out.
We also obtain, from theory as well as from MC simulations,
$m(t)$ for spatially disordered systems. More specifically,
we make a random selection of a fraction $\tilde n$ of
$L\times L\times L$ SC lattice sites and place spins on them.
For $\tilde n\lesssim 0.1$, we assume full
disorder. Then, theory predicts a magnetic relaxation
that bears no resemblance to a $\sqrt {t}$ rule, not even
to the $t^p$ rule that we obtain for fully occupied SC lattices.
Instead, for $\tilde n\lesssim 0.1$,
\begin{equation}
\tilde m\simeq \frac{80h_w}{\sigma}e^{-(t/\tau_m)^{q}},
\label{fit}
\end{equation}
approximately, where $\tau_m\simeq 10^6\Gamma^{-1}$ and $q\simeq -0.105$.
Results from our MC simulations are in fair agreement with this.
In Sect. \ref{tels} we report results for holes' line shapes in
fully occupied crystal lattices. 
The main results for line shapes, which are derived
in Sect. \ref{th}, follow.
When $1\ll \Gamma t\ll (\sigma/h_w)^{1/p}$ and
$h_w\ll \mid h+H\mid\ll \sigma$,
\begin{equation}
f(h,t)/f(h,0)\propto \mid \eta\mid^{\frac{1}{p}-1}
\label{eta2}
\end{equation}
if $\mid\eta\mid \ll 1$,
where 
\begin{equation}
\eta \equiv \frac{h+H}{h_w (\Gamma t)^{p}}, 
\label{etaeq} 
\end{equation}
$H$ is an applied field,
and $p$ is given by
\begin{equation}
p^{-1}\sin \pi p=\sqrt{2\pi}\sigma/h_0.
\label{pp}
\end{equation}
In Sect. \ref{th} we also derive relation for
holes line shapes that hold over longer time spans:
$1\lesssim \Gamma t\lesssim (\sigma /\varepsilon_w)^{1/p}$.
In Sect. \ref{cwe} we apply these results 
to published\cite{ww} experimental data 
for Fe$_8$ and to MC data for Fe$_8$ as well as to fully occupied
FCC lattices. Finally, 
concluding remarks appear in Sect. \ref{concl}.

\section{Relaxation of the magnetization}
\label{rm}

\subsection{Theory}
\label{rtr}

We first describe a stochastic model which helps to understand the physics
of the problem as well as the statistical assumption we have made
in order to solve it. Consider two tracks, both filled with particles.
Let there be one particle on the ``up-track'' for each up spin
on a lattice, and one particle in the ``down-track'' for each
down spin. Let all particles on each track be ordered according to the value
of the magnetic field $H+h$ acting on each spin. To mimic tunneling,
random select a particle within the tunnel window, that is, a particle
fulfilling $-h_w<H+h<h_w$, whether on the up- or down-track,
 and move it to the ``point''
$H+h$ on the opposite track. In order to mimic
the effect such a spin flip has on the dipolar fields
on other spins, draw a random value of $\Delta h$
for each particle from a Lorentzian distribution of half width $h_0/N$,
where $N$ is the total number of spins, and let $h\rightarrow h+\Delta h$
if the particle that just shifted track has done so for the first time.
This latter proviso is related to the fact that no effect on the dipolar
field follows when the same spin flips {\it twice} (for a more detailed
explanation, see Ref. [\onlinecite{ourPRB}]). Repeat this whole process
at every tic of a clock. Clearly, the whole process
stops when all particles haved jumped track at least once.

We can get a feeling for the relevance of lattice structure
or spatial spin distribution from the following simple
consideration. $Np(0)2h_w$ is the number of particles in
the tunnel window, which is the number of times the clock tics in
time $\Gamma^{-1}$, which, in turn, would be the relaxation time
for the equilibration of
the number of up and down particles in the tunneling window
if there were no ``field shifting''. Now, the median
field shift, or diffusion length, in time $\Gamma^{-1}$ is
$2p(0)h_0h_w$, that is, a fraction $p(0)h_0$ of the tunnel
window's width. Thus, $p(0)h_0$ is a measure of the
amount by which 
the unbalance
between up and down particles is restored
in the tunnel window in time $\Gamma^{-1}$. 
Therefore, the relaxation rate clearly depends on
$p(0)h_0$, which in turn depends on lattice structure. This shows
why the latter is relevant to the relaxation of the magnetization.

The simple statistical assumption above
has enabled us to derive\cite{ourPRB} the 
equations we need
for the calculation of the time evolution of the magnetization.
These equations, which we reproduce in a 
compact form immediately below, give the magnetization $m(t)$
at time $t$, and $n(t)$, the fractional number of spins that 
flip at least once in time $t$.
We first recall an important ingredient of
the theory for ZFC experiments:\cite{ourPRB}
the energy per spin at the time when the system is quenched,
which we refer to as the ``annealing energy'', is $-\varepsilon_a$.
Let $x_1=mg\mu_BS\sigma\langle h^2\rangle_0 /(\varepsilon_ah_wH)$ and
$x_1=2(\sigma /h_w)(1-m/m_0)$,
for ZFC and FC experiments (defined in the Introduction),
respectively, 
and $x_2=n\sigma /h_w$
for both FC and ZFC experiments.
The desired equation follows,
\begin{equation}
\frac{dx_j}{dt}\simeq a_j\sqrt{\frac{2}{\pi}}-b_j\int^t_0 d\tau
\frac{dx_j(\tau )}{d\tau}\frac{1}{\omega (t-\tau)+1},
\label{mfinal2}
\end{equation}
where
\begin{equation}
\omega(t^\prime )\simeq \min \left[\frac{\pi h_0}{2\sigma}x_2(t^\prime ),
\sqrt{\frac{\pi \sigma}{2h_w}x_2(t^\prime )}\right],
\label{omega}
\end{equation}
$a_1=4$, $b_1=2$.
In order to obtain
$x_1$, Eq. (\ref{mfinal2}) must first
be solved for $j=2$, letting $a_2=1$ and $b_2=1$, in order to then
use $x_2(t)$ in Eq. (\ref{omega}), and thus enable substitution of $\omega$ into
Eq. (\ref{mfinal2}) for $j=1$.

The theory applies if $h_w\ll\sigma$ and
the energy is constant, that is, if no energy
tranfer to the phonon bath takes place. 
This is also approximately so if
$kT\gg \varepsilon_w$. If the constant energy
condition is not met in a ZFC experiment, 
a linear in time magnetization
relaxation that is thermally driven takes over
before $m(t)\rightarrow m_s$.\cite{ourPRB}
The theory also gives
\begin{equation}
m_s\simeq 2H\varepsilon_a/(gS\mu_B\langle h^2\rangle_0). 
\label{ms}
\end{equation}

Note that the definition of $x_1(t)$, together
with Eqs. (\ref{mfinal2}),
(\ref{omega}), and (\ref{ms}) imply that the time
variation of $m(t)/m_s$ in a ZFC experiment
is the same
as $1-m/m_0$ in a FC experiment.

\begin{figure}[ht!]
\includegraphics*[width=80mm]{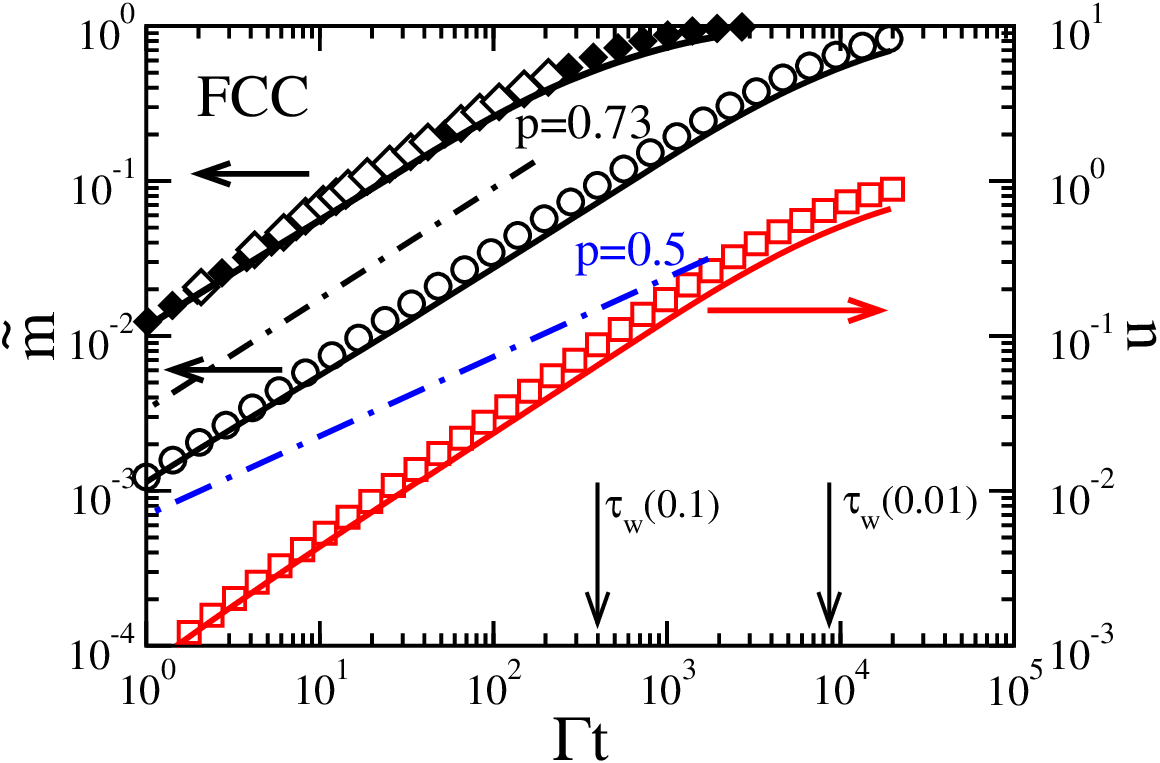}
\caption{$\tilde m$ and $n$ (the fractional number of spins that
have flipped in time $t$ at least once) versus time. 
Full lines are for theory and symbols stand for MC results for
$65536$ spins that are initially
up or down at random, with probabilities 
$0.6$ and $0.4$, respectively, on a FCC
lattice.
$\blacklozenge$ ($\circ$) stand for $\tilde m$ for
$h_w=0.1$ ($0.01$);
$\Box$ stand for $n$ for $h_w=0.01$.
The dash--dotted lines 
stand for slopes $p=0.73$,
and $p=0.5$, which follow for FCC lattices
from Eq. (\ref{pp}) and from
Ref. [\onlinecite{TSP}], respectively.
Data points taken from Ref. [\onlinecite{TSP2}]
are shown as $\Diamond$.
Our data points for $\Gamma t<1024$
come from averages over
$1400$ MC runs,
and for $100$ runs for $t>1200$.}
\label{magFCC}
\end{figure}

A few remarks about Eqs. (\ref{mfinal2}) and (\ref{omega})
are in order.
Clearly, $x_1(t)$ only depends on two parameters:
$h_0/\sigma$ and $\sigma /h_w$. The latter
only comes into play at the later portion
of the time evolution, when $n(t)> 2\sigma^2/(\pi h_0^2)$,
which is when $m(t)$ starts leveling off.
Now, $0.4\lesssim\sigma/h_0\leq\sqrt{\pi /2}$
for all cubic lattices, whether fully occupied
or not. It follows that leveling off of $m(t)$ is
triggered some time (see below) while $n(t)\leq 1$.
This has to do with the fact that a spin flip contributes to hole
widening in $f(h,t)$ the first time
it takes place after the field is switched on. When a spin
flips a second time, it only returns to its initial state,
thus {it cancelling} the effect os the first flip.
Since $m$ is proportional to
the width of the hole in $f(h,t)$ when $\Gamma t\gtrsim 1$
it follows that $m(t)$ becomes constant when $n\sim 1$.
The time when $m(t)$ levels off is illustrated in Fig. \ref{magFCC},
for fully occupied FCC lattices. For instance, for $h_w=0.01$,
theory predicts $m(t)$ to start leveling off when $n(t)\simeq 0.11$
since $2\sigma^2/(\pi h_0^2)\simeq 0.11$ then.
Thus, a significant portion of the evolution, up to $n\sim 1$, only depends on 
$h_0/\sigma$. Information about the lattice or
degree of spatial disorder only comes into the
equations through this number.
Other numbers,
such as $h_w$, $\varepsilon_a$,
and $H$ only come into the proportionality factor
between $x_1$ and either $m$ or $1-m/m_0$.
The temperature does not enter anywhere
into the equations.

We have shown analytically in Ref. [\onlinecite{ourPRB}]
that $m\propto t^p$ and $n\propto t^p$ fulfill Eq. (\ref{mfinal2})
for $j=1$ and $j=2$ in the 
$1\ll \Gamma t\ll (\sigma/h_w)^{1/p}$ time range
if $p$ is given by Eq. (\ref{pp}).
Numerical solutions of Eq. (\ref{mfinal2})
shows that $m\propto t^p$ ensues in the wider
$1\lesssim \Gamma t\lesssim (\sigma /h_w)^{1/p}$
time span for all fully occupied cubic lattices.\cite{ourPRB}
We show below that this is also so for a fully occupied
Fe$_8$ lattice. 

\begin{figure}[ht!]
\includegraphics*[width=80mm]{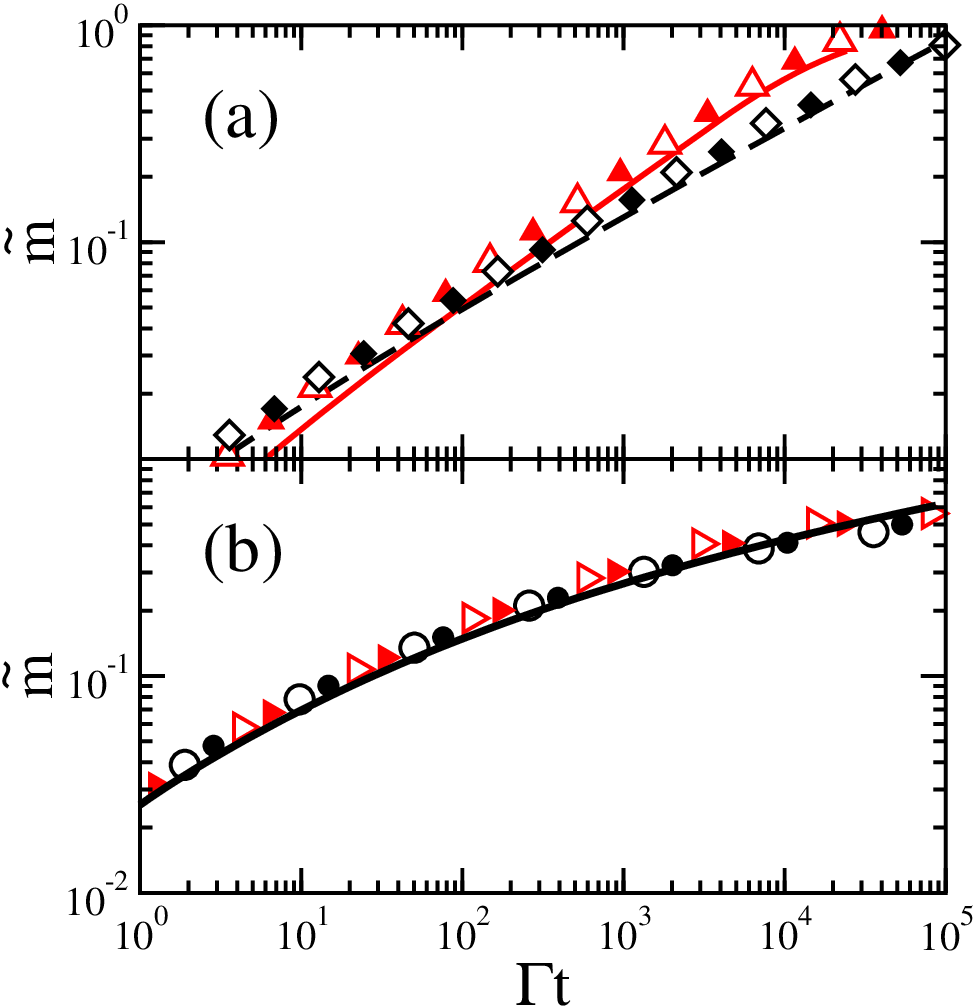}
\caption{\small
(a) Magnetization versus $\Gamma t$.
Lines are for theory and symbols are for data
from MC simulations of $L\times L\times L$ SC lattices
with a fraction $\tilde n$ of their sites occupied.
For $\tilde n=0.6$ the dashed line is for theory,
$\Diamond$, and $\blacklozenge$ are for MC data
for $L=32$ and $L=16$, respectively.
For $\tilde n=1$ the full line is for theory,
$\triangle$, and $\blacktriangle$ are for MC data
for $L=32$ and $L=16$, respectively.
All symbols are from averages of $800$ MC runs
while $\Gamma t<10^3$ and $40$ MC runs when $\Gamma t >10^3$.
(b) Same as in (a) but for $\tilde n=0.1$ and $h_w=0.02$,
and for $\tilde n=0.03$ and $h_w=0.006$.
The full line is for theory for both values of $\tilde n$.
For $\tilde n=0.03$ 
$\circ$, and $\bullet$ stand for MC data
for $L=64$ and $L=32$, respectively.
For $\tilde n=0.1$,
$\triangleright$, and $\blacktriangleright$ stand for MC data
for $L=32$ and $L=16$, respectively.
As in (a), all symbols are from averages of $800$ MC runs
while $\Gamma t<10^3$ and $40$ MC runs when $\Gamma t >10^3$.}
\label{magdil}
\end{figure}

Spatially disordered systems behave differently.
Consider a very small fraction ($\tilde n\ll 1$) of sites
of a SC lattice to be occuppied by randomly
oriented spins.
Then, a Lorentzian dipolar magnetic 
field distribution\cite{abrag} of half--width $h_0$ ensues.
It follows then, from the definition of $\sigma$
and of $h_0$, that
$\sigma /h_0=\sqrt{\pi /2}$. The number $p=0$
follows then from Eq. (\ref{pp}). Recall, however
that theory only implies that this ensues only when
$\Gamma t\gg 1$.
For earlier times, more specifically, for
all $\Gamma t\gtrsim 1$, we find that the numerical 
solution from Eq. (\ref{mfinal2})
is well fitted by Eq. (\ref{fit}) if $\tilde n\lesssim 0.1$. 
Numerical solutions of Eq. (\ref{mfinal2}) are
plotted in Fig. \ref{magdil}a for $\tilde n=1$ and
$\tilde n=0.6$, both for $h_w=0.02$ and in Fig. \ref{magdil}b
for (1) $\tilde n=0.1$ and $h_w=0.02$, and for (2)
$\tilde n=0.03$ and $h_w=0.006$. 
Note that the same solution obtains
for the latter two cases.
We come back to these
figures in Sect. \ref{1cwe}.

\subsection{Comparison with experiments and simulations}
\label{1cwe}

\begin{figure}[ht!]
\includegraphics*[width=80mm]{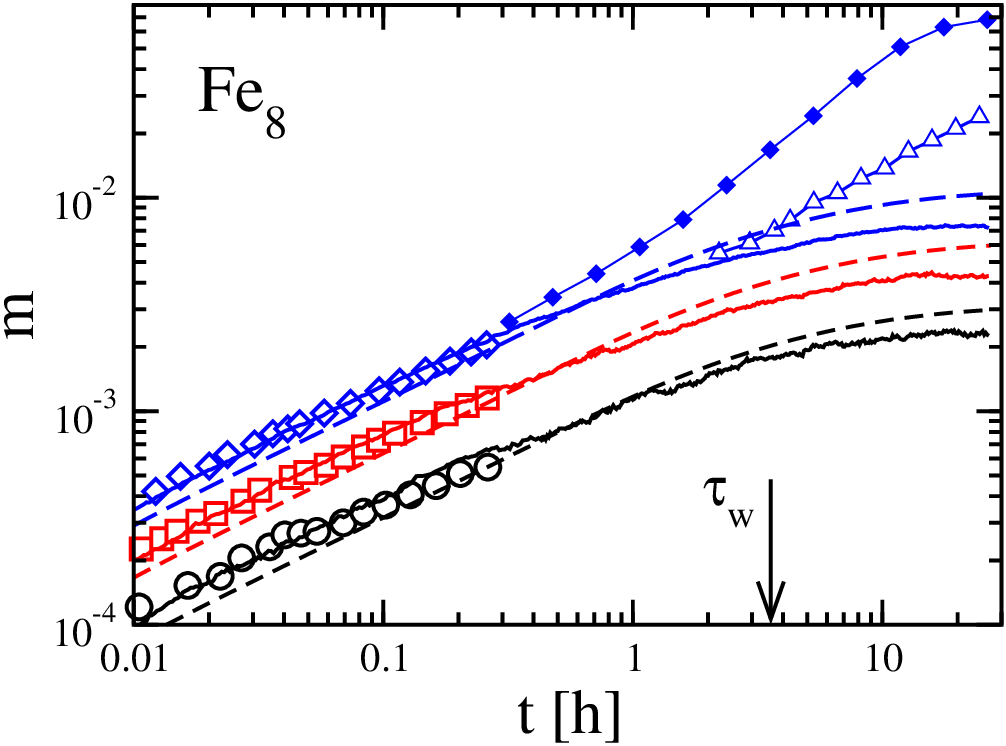}
\caption{\small
Magnetization versus time, in hours.
Units for $m$ are such that $m=1$ for full polarization.
$\Diamond$, $\Box$ and $\circ$
are for experimental data
taken from Ref.
[\onlinecite{ww}] for $H=3.92$, $2.24$, and $1.12$ mT, respectively.
Full lines are from our MC simulations for $kT\gg \varepsilon_w$,
and dashed lines are for theoretical predictions, that is,
from Eqs. (\ref{mfinal2}) and (\ref{omega}).
Data points that follow from MC simulations for $H=3.92$ mT
are also shown for $T=40$ ($\blacklozenge$) and $T=300$ mK
($\triangle$).
We assumed\cite{ww,ourPRB,JMM} $h_w=0.8$ mT and
used the values of $\sigma$ and $h_0$ that are given in Table I.
In the simulations, the initial state was 
prepared at $T=2$ K. At this temperature
we allowed the simulation to proceed in time  
up to the point when the energy of the system reached $-36$ mK,
which is $0.07$ of the ground state energy. This is the
value of $\varepsilon_a$ we used in order
to relate $x_1$ and $m$, just above Eq. (\ref{mfinal2}). We
treat one MC sweep as $\Gamma t=1$ and assume \cite{ourPRB}
$\Gamma =0.04$ s$^{-1}$ in order to convert MC sweeps to hours.
Note that $\tau_w\simeq 3 $ hours, since $\sigma\simeq 31$ mT
and $h_w\simeq 0.8$ mT. }
\label{mag}
\end{figure}

We first make use of Eqs. (\ref{mfinal2}) and (\ref{omega}) to
obtain $m(t)$ for Fe$_8$.
Some numbers must first be fed into
Eqs. (\ref{mfinal2}) and (\ref{omega}). For
$h_w$, we use\cite{mT} $0.8$ mT, as given in Ref. [\onlinecite{ww}].
We use $\Gamma =0.04$ s$^{-1}$, (see Refs. [\onlinecite{ourPRB}]
and [\onlinecite{JMM}]). 
With the numbers given in Table I for $\sigma$ and $h_0$, we obtain $x_1(t)$
and $x_2(t)$ numerically from Eqs. (\ref{mfinal2}) and (\ref{omega}).
Finally, the value of $-\varepsilon_a$, the annealing energy,\cite{ourPRB}
is needed in order to obtain $m$ from $x_1$. 
Not knowing $\varepsilon_a$, we treat it as a fitting
parameter. We find $\varepsilon_a\simeq 36$ mK fits
best the experimental data points
from Ref. [\onlinecite{ww}]), which are shown in Fig. \ref{mag}
for a few applied fields. The energy $- 36$ mK may
be compared to the approxiamte value, $-500$ mK, of the ground
state energy.\cite{err}

The MC data points shown in Fig. \ref{mag} follow
from simulations in which the system first evolves at some
high temperature (a few Kelvin) for a short time (less than 1 MC sweep)
until the energy equals $-36$ mK. At such temperatures,
Fe$_8$ cluster spins are not forced to tunnel through
the ground state doublet. Accordingly, all spins are
allowed to flip, regardless of the dipolar field
acting on them. We explore different scenarios
after quenching. In our theory, we assume no energy
exchange takes place between the spin system and
a heat reservoir. We have also performed MC simulations
under this assumption.
This is approximately realized for the time
range exhibited in Fig. (\ref{mag}) by flipping only spins
within the tunnel window, and then with equal probabilities
for upward or downward flips. If, on the other hand,
heat exchange does take place readily, as one gathers from Ref.
[\onlinecite{Mor}], where heat exchange rates that are
comparable to $\Gamma$ are found, then detailed balance
should be enforced in MC simulations. The results of doing this
lead to the plots shown in Fig. \ref{mag} for $T=40$ and $300$ mK.

Results obtained from theory, in Sect. \ref{rtr}
for very disordered systems are shown in Fig. \ref{magdil}b
for $\tilde n=0.1$ and $h_w=0.02$ and for $\tilde n=0.03$ and $h_w=0.006$.
Monte Carlo data points are also shown for the
same values of $\tilde n$ and of $h_w$.
Data points for $\tilde n=0.1$ and $h_w=0.02$
fall on top of data points for $\tilde n=0.03$ and  $h_w=0.006$.
This is as expected,
since $\sigma\propto \tilde n$, for
for full spatial disorder, implies that
$\sigma /h_w$ has the same value in both cases
and theory predicts independence from any other parameter.
The fitting function
from Eq. (\ref{fit}) falls right on top of
the curve for theory in Fig. \ref{magdil}b, and
cannot therefore be shown separately.

Finally, we consider size effects.
For a $16\times 16\times  16$ lattice and $\tilde n=0.1$,
for instance, approximately $410$ spins make up the system.
Of these, only approximately a fraction 
$2p(0)h_w$ are within the
tunnel window. That is, approximately $820h_w/\pi h_0$ spins,
which only amounts to some $12$ spins,
are within the tunnel window. For this reason, we
also simulated $32\times 32\times  32$ 
and $64\times 64\times 64$ lattices 
for $\tilde n=0.1$ and $\tilde n=0.03$ respectively. 
Monte Carlo data points are shown in Fig. \ref{magdil}.
Clearly, no significant size
effects are observed.
 
\section{Time evolution of the line shape}
\label{tels}

In this section we first derive some results for $f(h,t)$
that are valid whenever $m\propto t^p$ holds.
We know from Ref. [\onlinecite{ourPRB}] and from
the previous section that $m\propto t^p$ holds in
the time span $1\lesssim \Gamma t\lesssim (\sigma /h_w)^{1/p}$ 
for fully occupied SC, FCC, BCC, and Fe$_8$ lattices, but
we now know it is not so for 
spatially random systems. The results we derive below are
applied to fully occupied Fe$_8$ and FCC lattices and are 
compared to results from experiment
(for Fe$_8$) and from MC simulations. 

\subsection{Theory}
\label{th}

The starting point for the derivation 
are the following two equations, from Ref. [\onlinecite{ourPRB}],
\begin{equation}
g(h,t)\simeq \int_0^t d\tau \frac{d {m}(\tau 
)}{d\tau}G(h+H,t-\tau ) 
\label{ewe}
\end{equation}
and
\begin{equation}
G(h,H,t-\tau )\simeq \frac{u(t-\tau )}{\pi [(h+ H 
)^2+u(t-\tau )^2]},
\label{greenH}
\end{equation}
where $g(h,t)\equiv f(h,0)-f(h,t)$,
$u(t-\tau )\equiv h_0n(t-\tau )$, and $n(t-\tau )$
is the fractional number of spins that 
flip at least once in time $t-\tau$.
The rationale for these two equations is given next, but
(if $\mid h+H\mid\ll\sigma$ and $\Gamma t\ll (\sigma/h_w)^{1/p}$)
Eqs. (\ref {ewe}) and (\ref{greenH}) also follow 
from Eqs. (13)
and (14) of Ref. [\onlinecite{ourPRB}],
respectively.
Assume that, between times $t$ and $\tau$, a fraction $n(t-\tau )$
of all spins flip at least once and 
that $n(t-\tau )\ll 1$. This can later be checked
to be fulfilled if $t\ll \tau_w$, where 
$\tau_w \equiv \Gamma^{-1}(\sigma /h_w)^{1/p}$.  Then,
Eq. (\ref{greenH}) gives the probability density
$G(h,H,t-\tau )$ that, at time $t$, the field 
is $h+H$ at a site where
the field at time $\tau$ was $0$. \cite{abrag} 
To understand Eq. (\ref{ewe}),
note first that the definition of $g(h,t)$ implies 
that $g(h,t)$ must fulfill
$\int dh g(h,t)=m(t)-m(0)$. Equation (\ref{ewe}) does give 
$m(t)-m(0)=\int d\tau dm/d\tau$,
since $\int dh G(h,H,t-\tau )=1$. Similarly, 
a variation in the magnetization $(dm/d\tau )d\tau$ 
coming from some spin flipping between times $\tau$ 
and $\tau +d\tau$, when the field $h+H$
acting on them was within the tunnel window, contributes to
$g(h,t)$ with a Lorentzian curve whose width 
is $h_0n (t-\tau )$, where $n (t-\tau )/2$ is approximately the
fraction of the total number of sites where
spins at time $t$ point opposite to the way
they did at time $\tau$. \cite{ourPRB} Furthermore,
the area under the Lorentzian must be given by $(dm/d\tau )d\tau$.
That explains Eq. (\ref{ewe}).

\begin{figure}
\includegraphics*[width=80mm]{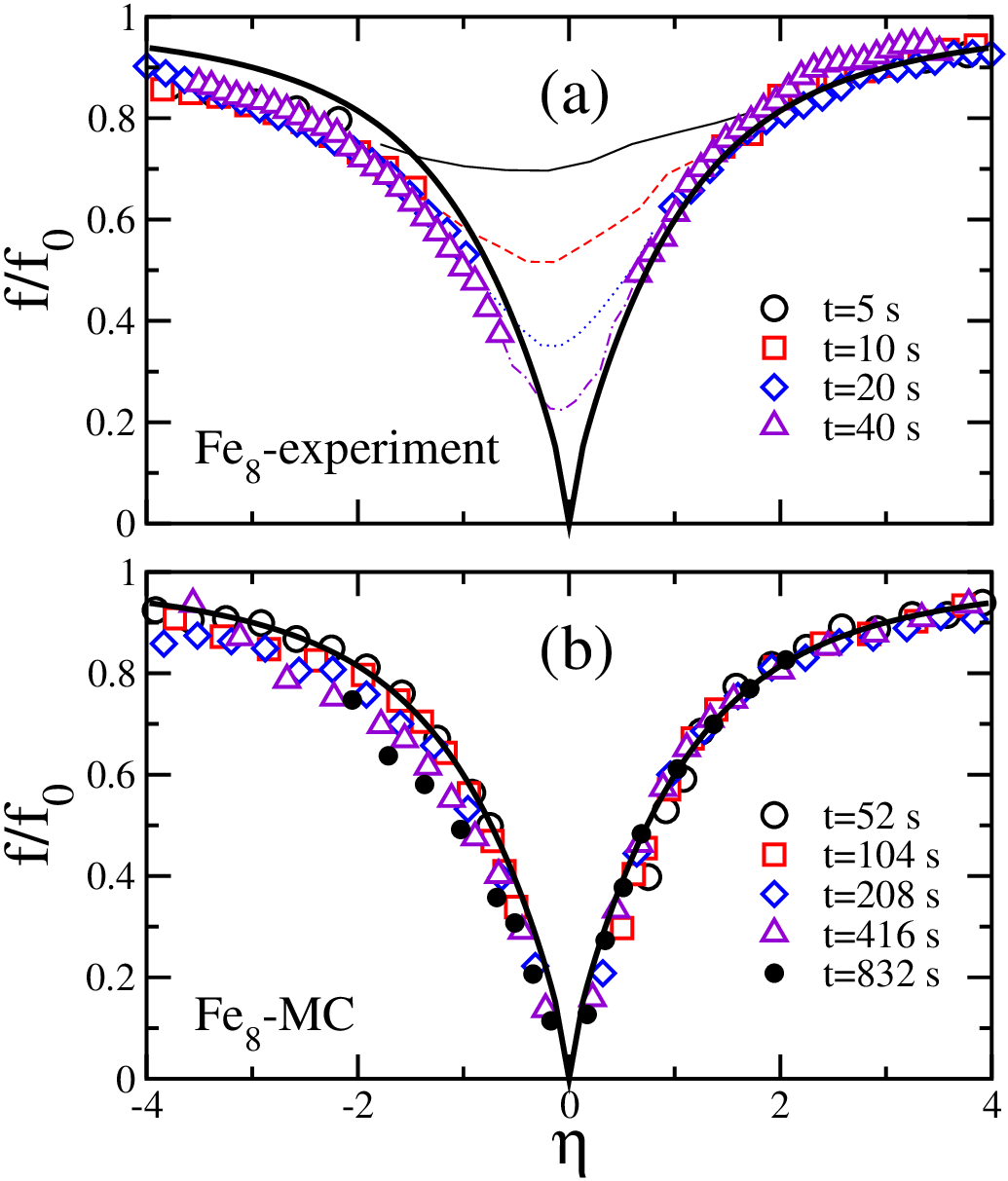}
\caption{(a) $f$ versus $\eta$, defined in Eq. (\ref{etaeq}),
for the times shown. All symbols stand for Wernsdorfer's
experimental data \cite{ww} for Fe$_8$
for $\mid h\mid >2h_w$.
Thin lines are for data points 
for $\mid h\mid <h_w$, to
which Eq. (\ref{eta2}) does not apply.
The full line
stands for Eq. (\ref{uno}), using $h_w=0.8$ mT
and $p=0.58$, from Eq. (\ref{pp}), and $\sigma /h_0=0.66$,
from Table I. (b)
Same as in (a) but for MC simulations of Fe$_8$.
Points from within the tunnel window have
been excluded.
Times are given in seconds, after converting each
MC sweep into $25$ seconds, which is the value we
assign to $\Gamma^{-1}$. All symbols stand for averages
over $1.6\times 10^5$ MC runs of systems of
$16\times 16\times 16$ spins.
The full line stands for Eq. (\ref{uno}).}
\label{exper}
\end{figure}

We make use of
$n(t-\tau )\simeq \tilde m(t-\tau )$,\cite{ourPRB} and of
\begin{equation}
\tilde m(\tau )\simeq 1.1 \frac{h_w}{\sigma}(\Gamma \tau )^p
\end{equation}
which has been shown to hold for all fully occupied
cubic and Fe$_8$ lattices in Ref. [\onlinecite{ourPRB}] and 
in Sect. \ref{1cwe}, respectively, while
$1\lesssim (\Gamma \tau )^p\lesssim \sigma /h_w$.
We then divide by $f(h,0)$, which, for $h\ll \sigma$ is 
approximately given by $m_0/\sqrt{2\pi}\sigma$.\cite{ourPRB}
Finally, the change of variable
$x\equiv \tau /t$
brings, if
$1 \ll \Gamma t\ll (\sigma/h_w)^{1/p}$ and
$h_w\ll \mid h+H\mid\ll \sigma$, Eq. ({\ref{ewe}) into
\begin{equation}
\frac{f(h,t)}{f(h,0)}\simeq
1-\frac{\sin \pi p}{\pi}\int_0^1 dx x^{p-1}\frac{(1-x)^p}
{\alpha\eta^2+(1-x)^{2p}},
\label{uno}
\end{equation}
where $\alpha\simeq 0.8(\sigma/ h_0)^2$.
Both Eqs. (\ref{eta2}) and (\ref{uno}) show that
the field $h$ scales as $t^p$ in holes' line shapes.

Finally, to obtain Eq. (\ref{eta2}) from Eq. (\ref{uno}),
note first that $f(0,t)/f(0,0)\rightarrow 0$ as 
$t\rightarrow \infty$ in Eq. (\ref{uno}),
since the integral therein equals 
$\pi /\sin p\pi$ if $\eta =0$. \cite{tablas}
Then, breaking up the integration interval into two pieces,
(1) from $0$ to $(\alpha\eta^2)^{1/2p}$, and (2) from $(\alpha\eta^2)^{1/2p}$
to $1$, and expanding $x^p/(\alpha\eta^2+x^{2p})$ in powers of $\epsilon$
and $1/\epsilon$ in the first and second integration intervals, respectively,
where $\epsilon\equiv x^{2p}/\alpha\eta^2$, gives $\mid \eta\mid ^{1/p-1}$ 
for the leading term, which is desired result,
that is, Eq. (\ref{eta2}).

\subsection{Comparison with experiments and simulations}
\label{cwe}

In this section we test our results, that is, the validity of
Eqs. (\ref{eta2}) and (\ref{uno}),
against experiments\cite{ww} and against
our MC simulations.

\begin{figure}[ht!]
\includegraphics*[width=80mm]{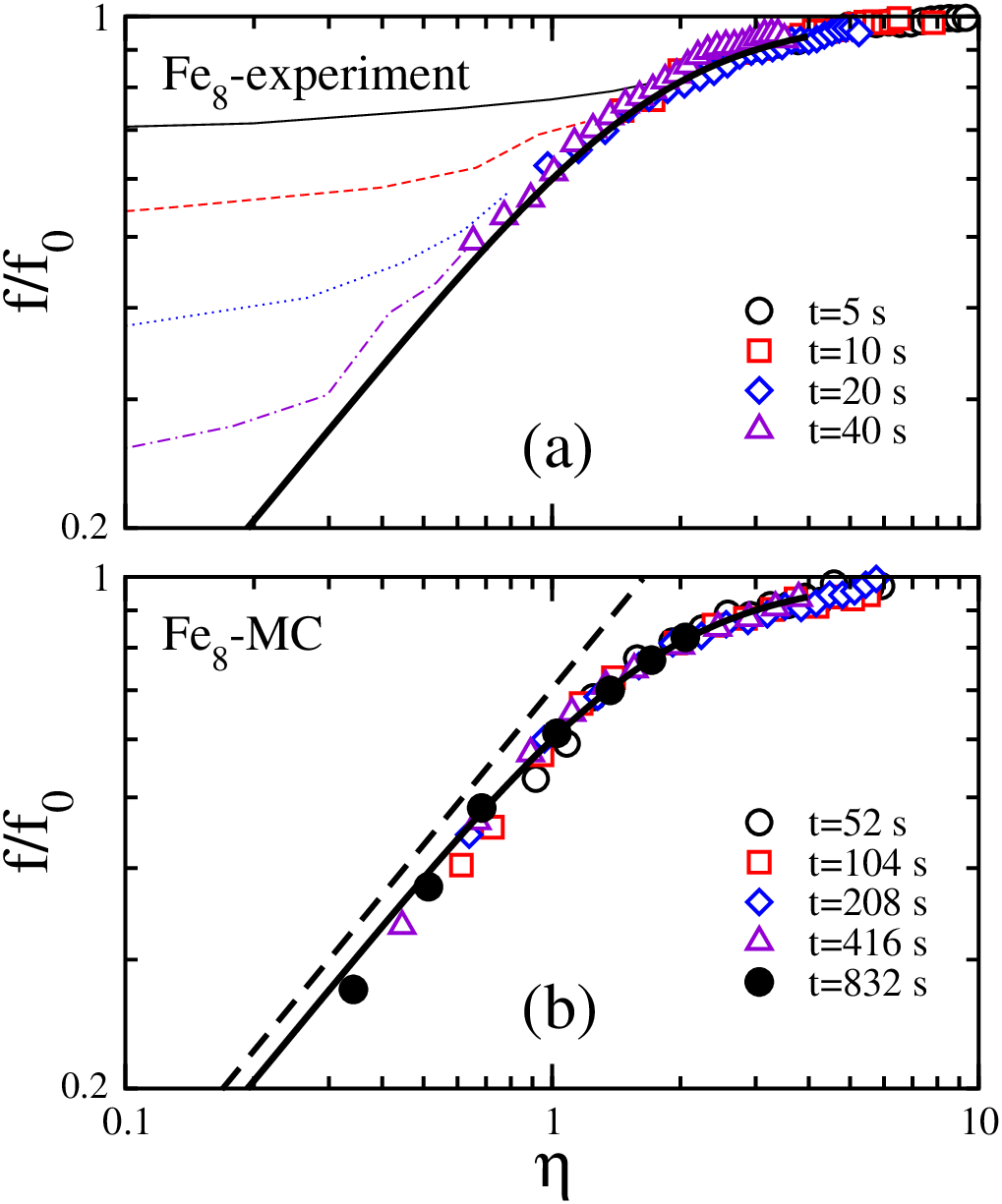}
\caption{(a)Everything is as in Fig. \ref{exper}a,
except that a Log--log scale is used here.
(b) Same as in (a) but for MC simulations, instead
of experiments.  Everything else is as in Fig. \ref{exper}b.
The dashed line stands for $\eta^{1/p-1}$,
as predicted by Eq. (\ref{eta2}) for $p=0.58$, for Fe$_8$.}
\label{experlog}
\end{figure}

We first apply Eqs. (\ref{eta2}) and (\ref{uno})
to Fe$_8$. From Table I,
$\sigma /h_0=0.66$ follows. Substitution of this number into
Eq. (\ref{pp}) gives
$p=0.58$.
Knowing the value of $p$ enables us to plot the data points for 
Fe$_8$ 
shown in Fig. \ref{exper}a.
Unfortunately, 
data for holes in Fe$_8$ have only been published
for $t\leq 40$ s, that is, for $\Gamma t\lesssim 1.6$, a
time which falls short of the validity range 
for Eqs. (\ref{eta2}) and (\ref{uno}). 
Still, one can appreciate in Fig. \ref{exper}a
how the data points seem to approach the theory curve 
for $f(h,t)/f(h,0)$ as $t$
increases up to $\Gamma t\lesssim 1.6$. 

In order to see how this would go for longer times, we have
used our model to simulate an experiment on Fe$_8$. 
The results are shown in
Fig. \ref{exper}b. We have let one MC sweep equal
$\Gamma t=1$, which, by the argument given above, 
implies $t\simeq 25$ s for Fe$_8$.
The agreement with theory is remarkable. This is better appreciated
in the log--log plot shown, with the same data, in Figs. \ref{experlog}a and b.
On the other hand, rescaling these plots, using $p=1/2$
gives rise to some data point scatter, but
not sufficiently large to convincingly rule out $p=1/2$. This is
not too surprising, given the small difference between $p=1/2$ and
the value $p=0.58$ that is given by Eq. (\ref{pp}). Still, one might have
hoped that these data would have been sufficient to discriminate
between Eq. (\ref{eta2}), where $\eta$ is raised to the $1/p-1$
power and the Lorentzian curve of Ref. [\onlinecite{TSP}].
Again, data for smaller values of $\eta$ would be required for this.
We know of no other
experimental results for hole digging we can make use of. 
As far as we know, all other reported experiments 
for SMM systems start from strongly polarized initial states.
\cite{info} 

Consequently, we decided to do simulations of SMM's in
FCC lattices, because Eq. (\ref{pp}) 
gives then a value, $0.73$, for $p$,
which differs significantly from $1/2$. We are now at liberty to
choose the value of $h_w$. In order to be able to obtain
holes' line shapes 
down to rather small values of $\eta$, and still meet 
the validity criterion for Eqs. (\ref{eta2}) and (\ref{uno}),
 we let $h_w$ take values down to $0.01$.

We show how $m$ varies with $t$ in Fig. \ref{magFCC} 
in a simulated experiment in which all spins are up and
down with probabilities $0.6$ and $0.4$ in the initial state,
and evolve thereafter with no applied field.
For $\Gamma t\lesssim 1$ (not shown),
$\tilde m\propto \Gamma t$. Note that $\tilde m\propto (\Gamma t)^p$ up to
$\Gamma t\sim (\sigma /h_w)^{1/p}$, which for $h_w=0.01$, for instance,
$\Gamma t\sim 9\times 10^3$, as predicted.\cite{ourPRB}
Note also the good agreeement with the value, $0.73$, 
which Eq. (\ref{pp}) gives for $p$.

\begin{figure}[ht!]
\includegraphics*[width=80mm]{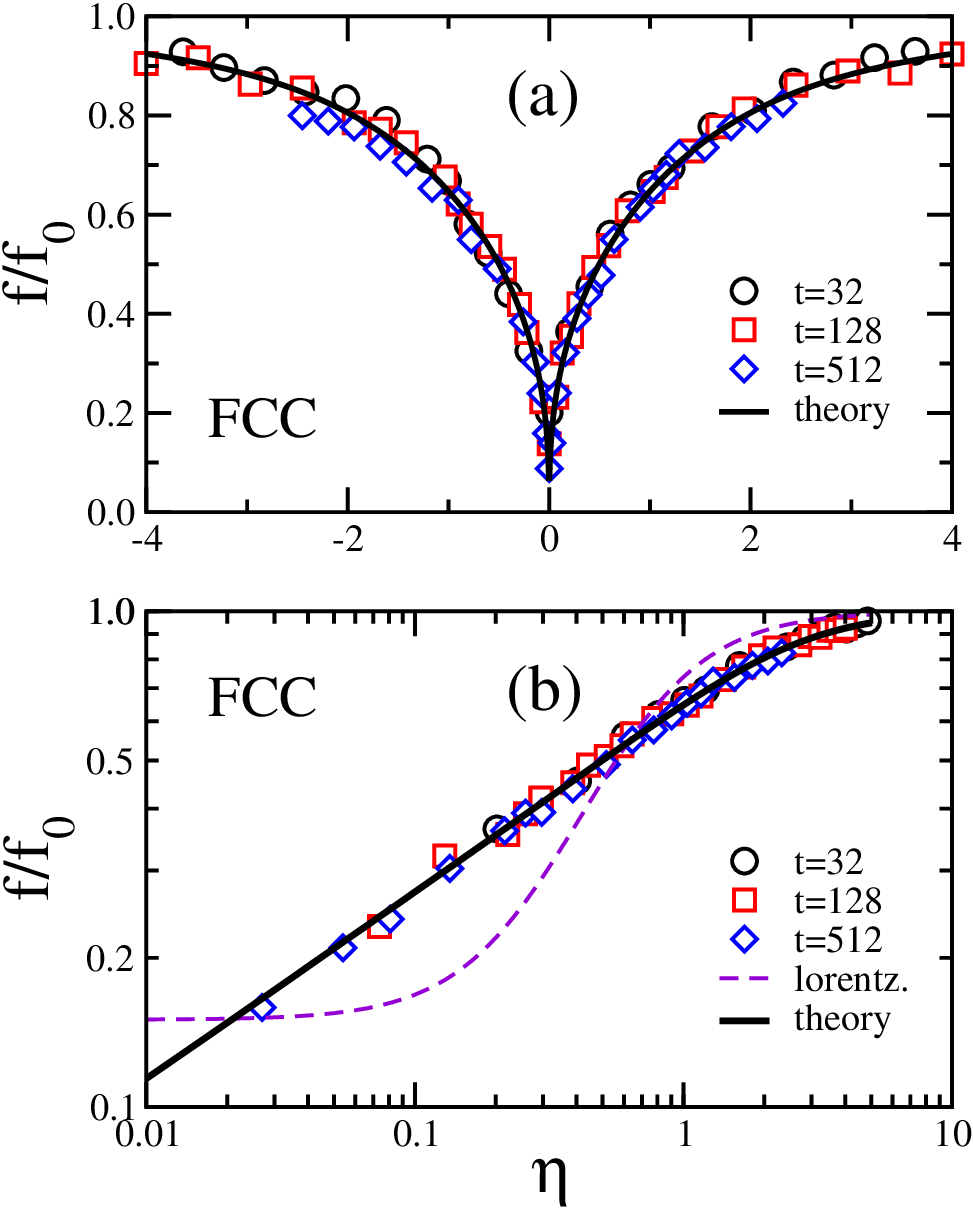}
\caption{(a) $f(h,t)/f(h,0)$ versus $\eta$, defined in Eq. (\ref{etaeq}),
for the shown times. Times are in MC sweeps.
$h_w=0.01$.
Symbols stand for averages over $1400$ MC runs
for $65536$ spins in a FCC lattice.
The full line
is from Eq. (\ref{uno}). As in Fig. 4, initially, all
$65536$ spins are randomly
up or down with probabilities $0.6$ and $0.4$, respectively.
(b) Same as in (a) but in a Log--Log scale.
The solid line stands for Eq. (\ref{uno})
and the dashed line is for the best fitting Lorentzian curve.
The dash--dot line stands for $\eta^{1/p-1}$,
as predicted by Eq. (\ref{eta2}), for $p=0.73$, given
by Eq. (\ref{pp}) for FCC lattices.}
\label{FCC}
\end{figure}

Holes' line shapes obtained from MC simulations are shown
in Figs. \ref{FCC}a and b. The nice 
agreement with theory is reassuring.
Similar plots but using $p=0.5$
give unsatisfactorily wide data point scatter.
The data clearly follow Eq. (\ref{eta2}) for $\eta\ll 1$ and 
deviate sharply from a Lorentzian line shape.
We do not exhibit results for SC or BCC lattices,
but we have found them to follow
our predictions equally well.

\section{conclusions}
\label{concl}

Results that follow from our theory for the relaxation
of the magnetization of interacting SMMs are reported.
They are in fair agreement with the experimental
relaxation of the magnetization observed in Fe$_8$
as well as with our own MC results for Fe$_8$ and
for other lattices.
Furthermore, we make some predictions for Fe$_8$ that
can be checked experimentally.
Experiments following the lead of Ref. [\onlinecite{ww}]
would be made feasible by the application
of a transverse field $H_\perp$ of approximately $ 0.3$ T,
since $\Gamma$, which increases as $\Delta ^2$,
would then increase by a factor of roughly $50$
if $H_\perp$ is applied along the easier magnetization direction on the
$xy$ plane
(see Figs. 2 and 3 of Ref. [\onlinecite{sc}]).
This would in effect approximately reduce the time scale in Fig. \ref{mag}
from hours to minutes. 
Comparison of experimental results with MC data
shown in Fig. \ref{mag} would be interesting.
It would, for instance, show whether
heat exchange takes place readily, as inferred in Ref. [\onlinecite{Mor}]
from nuclear magnetic resonance experiments, or not.

We have also shown, from theory and MC simulations,
that the magnetization of 
spatially disordered SMMs
relaxes, not as any power of time,
but approximately as given by Eq. (\ref{fit}).
A counterintuitive 
prediction that follows from our theory and from
MC simulations can be gathered from
Figs. \ref{magdil}a and \ref{magdil}b. One might have thought
that dilution would lead to weaker dipole interactions
and, consequently, to unhindered, faster relaxation.
Instead, the opposite effect takes place for 
$0.1\lesssim \tilde n\leq 1$ after some time.

Line shapes
that develop in crystals of Fe$_8$ clusters
have been obtained from our theory. We have
shown that $f(h,t)$ is only a function of $h/t^p$
for all $\mid h\mid >h_w$, that is, for all $h$ outside
the tunnel window.
This is the main content of  Eq. (\ref{uno}). 
Furthermore, we have shown that
data points from experiments on Fe$_8$, taken from 
Ref. [\onlinecite{ww}], as well as results from MC simulations we have
performed for the same system, follow this rule. 
Scaling also ensues
for the data from our MC 
simulations of SMM's on FCC
lattices for $p=0.73$,
as given by Eq. (\ref{pp}), but not for
the otherwise predicted\cite{TSP,comm,TSP2} $p=1/2$ value that is 
supposed to hold universally. 

We have also shown that $f(h,t)\sim \mid h/t^p\mid ^{(1/p)-1}$ 
if $h_w/\sigma\ll \eta\ll 1$ and 
$h$ is outside the tunnel window.
Again, this is in agreement 
with experimental and MC results for Fe$_8$ 
(see Figs. \ref{experlog}a and b), FCC 
(see Figs. \ref{FCC}a and b), and (not shown) SC and BCC lattices.
A rough argument that explains why holes'
line shapes are not Lorentzian follows.
Note first that while field distributions
from dilute systems of dipoles are indeed Lorentzian,\cite{abrag}
only spin flips that take place {\it after} time $t$ contribute to
the diffusion of a hole that was ``dug'' at time $t$.
Since a full hole is only dug gradually in the course of time,
a {\it sum} of Lorentzian functions of $h$ [see Eq. (\ref{uno})] of various
widths is expected. Not surprisingly, Lorentzian function
does not ensue for $f(h,t)$ [see, Eq. (\ref{eta2})].
Here, experimental data for holes in the hundreds of seconds
time range, over which the magnetization has already 
been observed experimentally,
would be helpful.

Financial support from grant BFM2003-03919/FISI, 
from the Ministerio de Ciencia y 
Tecnolog\'{\i}a of Spain, is gratefully acknowledged.


\begin{thebibliography}{9}
\bibitem{PS2}Except perhaps for an erroneous 
statement that
is made in Ref. [\onlinecite{PS}], to the effect that relaxation
from weakly polarized states proceeds exponentially.
\bibitem{more}MC results for $FCC$ lattices that
are in sharp disagreement with the
scaling rule $f(h,t)=f(h/t^2)$, which TSP hold to
be lattice independent are exhibited in Ref. [\onlinecite{comm}].
\bibitem{ncc}More generally, for non-cubic lattices, 
$h_0=(8\pi^2/3^{5/2})g\mu_BS\rho$,
where $\rho$ is the dipole number density.
\bibitem{mT}Whether $h_w\simeq 0.8$ mT as it appears in Ref.
[\onlinecite{ww}], or $h_w\simeq 0.4$ mT for ``standard'' Fe$_8$,
as it appears in [\onlinecite{wwq}] is not too important
for our purposes. If $h_w\simeq 0.4$, then $\varepsilon_w\simeq -72$ mT,
instead of $\varepsilon_w\simeq -36$ mT,
would have to be assumed, since only the product $h_w\varepsilon_w$
enters into the theory.
\end{thebibliography}
\end{document}